\documentclass[aps,twocolumn,showpacs,groupedaddress,nofootinbib]{revtex4}
\usepackage{epsfig}
\usepackage{bm}

\newcommand{\Jetset}{\textsc{jetset}}

\begin{document}

\preprint{02-02}

\title{Pionic transparency in semi-exclusive electroproduction off nuclei}
\author{Murat M. Kaskulov}
\email{kaskulov@theo.physik.uni-giessen.de}
\author{Kai Gallmeister}
\author{Ulrich Mosel}
\affiliation{Institut f\"ur Theoretische Physik, Universit\"at Giessen,
             Germany}
\date{\today}

\begin{abstract}
We investigate the early onset of pionic color transparency ($\pi$CT)  observed
at Jefferson Laboratory (JLAB) in semi--exclusive pion electroproduction
reaction  $A(e,e'\pi^+)$ off nuclei. In the present description the primary
$\gamma^*p \to \pi^+ n$ interaction is described very well both for the longitudinal and the
transverse  polarizations. For the final state interactions a coupled--channel
treatment of the interactions of transmitted hadrons 
allows to go beyond the Glauber approximation.  
We show that a proper distinction between  the soft hadronic and hard
partonic components of the electroproduction amplitude is essential for a
quantitative description of the measured nuclear transparency.  
The data are well reproduced if one assumes that point--like configurations are
produced in the regime of hard deep--inelastic scattering (DIS) off partons and
dominate the transverse channel.
\end{abstract}
\pacs{25.20.Lj, 13.75.Gx, 12.39.Fe, 11.80.La, 13.40.Gp}
\maketitle

%\special{!userdict begin /bop-hook{gsave 200 30
%translate 65 rotate
%/Times-Roman findfont 240 scalefont setfont 0 0 moveto
%0.90 setgray (DRAFT)
%show grestore}def end}

The interactions of high--energy virtual photons with nuclei
provide an important tool to study the early stage of hadronization
and (pre)hadronic final--state--interactions (FSI)
at small distances $d \sim 1/\sqrt{Q^2}$. % from the origin.
A further advantage of lepton--induced reactions is that one
may vary the energy $\nu$ and virtuality $Q^2$ of the incident photon 
independently of each other. This allows to study the phenomenon of Color
Transparency (CT)~\cite{Mueller:1982,Jain:1995dd}, i.e. the reduced
interaction cross section of a small sized color singlet object produced in
processes at high momentum transfer. Furthermore, in the kinematic regime 
where one is
less sensitive to the resolved hadronic interactions of photons
-- coherence length effect -- the photonuclear reactions are not contaminated by
initial--state--interactions (ISI). This is an advantage compared to the use
of hadronic projectiles, which are strongly shadowed on their way to the
reaction point inside the nucleus.

In the presence of the CT effect the intranuclear attenuation of hadrons
propagating through the nuclear medium should decrease as a function of photon
virtuality $Q^2$. In this case the nucleus becomes more transparent for the outgoing
particles as compared to the case where the attenuation is driven by
ordinary absorption mechanisms. One may use a transparency
ratio\footnote{The nuclear transparency for a certain reaction process is usually defined
as the ratio of the nuclear cross section per target nucleon to the one for a
free nucleon, i.e. $T_A=\sigma_A/A\sigma_N$.} as a tool to search for
deviations from predictions of models based on conventional nuclear many--body mechanisms.

During the last decade, a number of experiments have been performed to measure
the nuclear transparencies in search of CT. These include the measurements
of transparencies in reactions $A(p,2p)$~\cite{Carroll:1988rp,Mardor:1998,
Leksanov:2001ui,Aclander:2004zm}, $A(e,e'p)$~\cite{Garino:1992ca,Makins:1994mm,
O'Neill:1994mg, Abbott:1997bc,Garrow:2001di,Dutta:2003yt}, $\rho$--meson 
electroproduction off nuclei~\cite{Adams:1994bw,Airapetian:2002eh},
$\pi$--photoproduction~\cite{Dutta:2003mk}, coherent diffractive dissociation
of pions into di--jets~\cite{Aitala:2000hc}.  We refer to
Ref.~\cite{Strikman:2007nv}, where a summary of the
possible signatures of CT in these reactions can be found.

It is expected that CT should be more pronounced in reactions involving
mesons instead of baryons. Indeed, it might appear more probable to produce a two--quark
system with a small transverse size and, as a consequence, reduced FSI.
This idea had been followed by the HERMES experiment on $\rho$--meson
production on nuclei \cite{Airapetian:2002eh} which seemed to indicate
CT effects. Subsequent theoretical analyses of this experiment did show,
however, that the effects of CT could not be clearly distinguished from
those of shadowing in the entrance channel
\cite{Kopeliovich:2001xj,Falter:2002vr}. The latter calculation
\cite{Falter:2002vr} showed that it is essential in any analysis of the
CT effect to properly account for FSI and experimental
acceptance limitations. The coherence length has been kept approximately
constant in an experiment at JLAB~\cite{:2007gqa} where the nuclear
transparency in semi--exclusive $\pi^+$ electroproduction reaction
$A(e,e'\pi^+)$ has been measured as a function of $Q^2$ and the atomic mass 
number $A$, a rise of pionic transparency has been observed for values of $Q^2$
between 1 and 5 GeV$^2$. Since the $\pi$--nucleon cross section is nearly
constant for energies covered by the experiment, transparencies calculated
in the standard Glauber model %~\cite{Glauber} 
are independent of $Q^2$. The observed $Q^2$ dependence of the $\pi^+$
transparency deviates from the calculations without CT of
Refs.~\cite{Larson06ge,Cosyn:2007er}, and is in  fair agreement with
the calculations of the same groups when including CT. In Ref.~\cite{Larson06ge}
a semiclassical model for FSI has been used, while in Ref.~\cite{Cosyn:2007er}
a relativistic version of the Glauber model
has been developed. Both groups incorporate CT using the
quantum diffusion picture~\cite{farrar}.

In this work we study the onset of CT at JLAB
using a factorization of the whole reaction into an initial, primary
interaction of the incoming virtual photon with the nucleon and the FSI.
The cross section for the former is reproduced both in its longitudinal and
its transverse contribution while the FSI is treated within
transport approach. The main advantage of the present approach to FSI is its
universality and a use of input parameters already tested in many different
kinds of  nuclear reactions,  for example, for photo-- and electroproduction
reactions~\cite{Buss:2007ar}. For the present investigation the studies of
pion--processes in nuclei are particularly relevant~\cite{Buss:2007sa,Krusche:2004zc}.
Because our calculations represent a complete event simulation it is possible 
to take experimental acceptance effects  into account. In the following we carry out
the transport MC simulation using the actual
acceptance conditions of the  $\pi$CT experiment at
JLAB~\cite{:2007gqa,Clasie:2006re}.  As a central result we will show that a
quantitative understanding of the observed nuclear transparency requires a
detailed understanding of the primary, elementary $(\gamma^*,\pi^+)$ reaction on the proton.
In spite of the fact that the results of Refs.~\cite{Larson06ge,Cosyn:2007er}
do provide a rather strong support for the CT 
we further show that because of large uncertainties in the
formation time concepts the longitudinal--transverse ({\textsc
l}--{\textsc t}) separated nuclear cross sections are needed 
for a quantitative understanding and proof of the CT effect.

\begin{figure}[t]
\begin{center}
\includegraphics[clip=true,width=1\columnwidth,angle=0.]{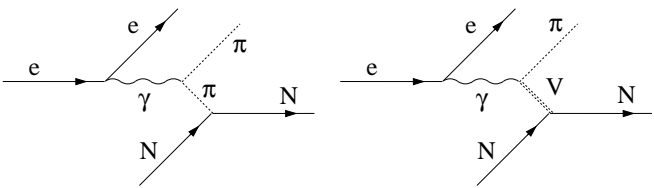}
\caption{\label{Figure1} 
\small The diagrams describing the hadron--exchange part of
       the $\pi^+$-- electroproduction amplitude. See text for
       the details.
\vspace{-0.7cm}
}
\end{center}
\end{figure}

\begin{figure}[b]
\begin{center}
\includegraphics[clip=true,width=0.95\columnwidth,angle=0.]
{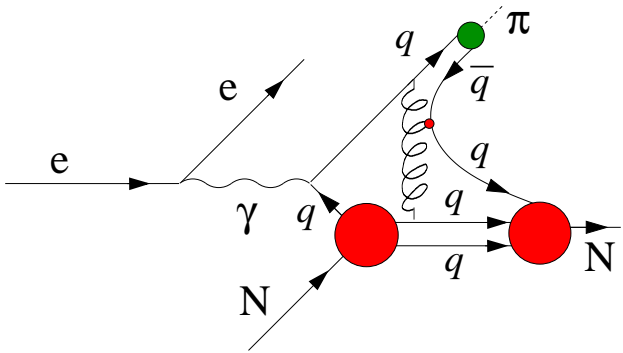}
\caption{\label{Figure2}
\small A schematic representation of the partonic DIS part of the $\pi^+$--
electroproduction mechanism. The wavy line represents a color string.
See text for the details.
\vspace{-0.7cm}
}
\end{center}
\end{figure}

\begin{figure}[t]
\begin{center}
\includegraphics[clip=true,width=0.9\columnwidth,angle=0.]
{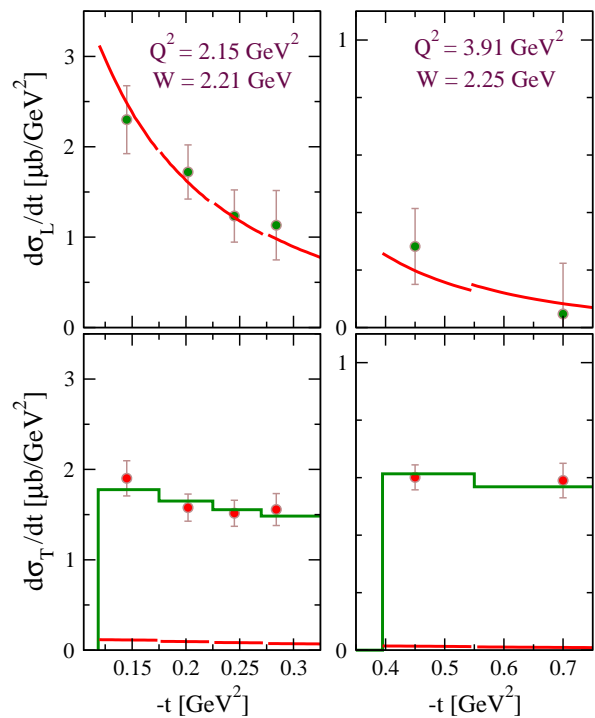}
\caption{\label{Figure3}
The longitudinal $d\sigma_{\rm L}/dt$~(top panels) and
transverse $d\sigma_{\rm T}/dt$ (bottom panels) differential cross sections
of the reaction $p(e,e'\pi^+)n$ at average values of
$Q^2=2.15$~GeV$^2$ and $Q^2=3.91$~GeV$^2$.
The solid curves are the contribution of the hadron--exchange model
and the histograms are the contribution of the DIS pions.
The $\pi$CT data are from Ref.~\cite{Horn:2007ug}.
The discontinuities in the curves result from
the different values of $Q^2$ and $W$ for the various $-t$ bins.
\vspace{-0.7cm}
}
\end{center}
\end{figure}

The calculations of Refs.~\cite{Larson06ge,Cosyn:2007er} mainly deal with
effects of FSI and do not account for many features of the $(\gamma^*,\pi^+)$
reaction amplitude. Our interpretation of the $\pi$CT data is closely tied to
a microscopic model for the primary reaction ${p}(e,e'\pi^+){n}$. At first we
briefly describe the model for the exclusive process
\begin{equation}
\gamma^{*}(q) + N(p) \to \pi(k') + N'(p').
\end{equation}
Following Ref.~\cite{Kaskulov:2008xc} we distinguish two classes of primary
collisions: a soft hadronic and a hard partonic (DIS) production of $\pi^+$.
The soft hadron--exchange part of the $\gamma^* p \to n \pi^+$ amplitude is
described by the exchange of Regge trajectories~\cite{Vanderhaeghen:1997ts}.
The Feynman diagrams are shown in Figure~\ref{Figure1}. The left diagram in
Figure~\ref{Figure1} corresponds to the exchange of the $\pi$--Regge
trajectory and is referred as the $\pi^+$ quasi--elastic knockout mechanism.
The latter also contains the electric part of the $s$--channel
nucleon Born term to conserve the charge of the system.
The right diagram in Figure~\ref{Figure1} describes the exchange
of the $\rho$--meson Regge trajectory and gives only a marginal contribution
to the cross section. The corresponding gauge invariant
hadronic currents and parameters of the Regge trajectories are given in 
Ref.~\cite{Kaskulov:2008xc}.

At the  invariant masses reached in the $\pi$CT experiment ($W
\approx 2.2$ GeV)  nucleon resonances can contribute to the $1\pi$ channel.
As in Ref.~\cite{Kaskulov:2008xc} this is modeled by the hard interaction
of virtual photons with partons (DIS)  since  DIS involves all possible
transitions of the nucleon from its ground state to any excited state~\cite{Close:2001ha}.
For the proper description of the reaction $p(e,e'\pi^+)n$ in DIS 
a model for the hadronization process is needed. In the present 
description of the hadronization in DIS we rely on the Lund fragmentation
model~\cite{Andersson:1983ia} 
as depicted in Figure~\ref{Figure2}
where the leading order $\gamma^* q \to q$ DIS process followed by the 
fragmentation of an excited string into two particles ($\pi N$)
is shown. As a
realization of the Lund model we use the \Jetset{} implementation~\cite{Sjostrand:2006za}.
We note that this description resolves the longstanding puzzle of a
large theoretical underestimate of the observed transverse strength in
the model of~\cite{Vanderhaeghen:1997ts}.

In the calculation presented in Ref.~\cite{Kaskulov:2008xc} 
the transverse part is solely generated by the DIS process. 
However, contrary to the situation at higher values of $Q^2$ 
considered in Ref.~\cite{Kaskulov:2008xc}
where the 
hadronic part gives only a marginal
contribution to $\sigma_{\rm T}$, at low $Q^2$ 
the problem of double counting arises
when using both the DIS and the Regge contributions to the transverse cross
section. Following Ref.~\cite{Friberg:2000nx} 
this could be solved by turning off the leading order  DIS
contribution, as required by gauge
invariance for $\gamma^* q \to q$, when approaching the photon point 
where the Regge description
alone gives a good description of data~\cite{Vanderhaeghen:1997ts}. 
Therefore, an additional empirical factor~\cite{Friberg:2000nx} 
\begin{equation}
\label{DIScf}
{Q^4}/{(Q^2 + \Lambda^2)^2}
\end{equation}
is introduced into the DIS cross section with the cut--off $\Lambda$ as a fit
parameter. This factor is $\simeq 1$ at high values of $Q^2$ and tends to zero at small values of
$Q^2$. The combined description of data from the JLAB $F\pi$--1 experiment~\cite{Tadevosyan:2007yd}
at low values of $Q^2$ and $F\pi$--2 experiment~\cite{Horn:2006tm}
at high $Q^2$ results in $\Lambda \simeq 400$~MeV. In the $Q^2$ range
considered in Ref.~\cite{Kaskulov:2008xc} Eq.~(\ref{DIScf}) is close to unity
and is largely ineffective for the results presented there.

In Figure~\ref{Figure3} the results for the $p(\gamma^*,\pi^+)n$ differential
cross sections $d\sigma_{\rm L}/dt$ (top panels) and $d\sigma_{\rm
  T}/dt$~(bottom panels) are compared with the $\pi$CT data of
Ref.~\cite{Horn:2007ug}. The longitudinal cross section $d\sigma_{\rm L}/dt$
is very well described by the hadron--exchange model (solid curves). The
discontinuities in the curves result from the different values of $Q^2$ and
$W$ for the various $-t$ bins. The steep fall of $d\sigma_{\rm L}/dt$ away
from forward angles comes entirely from the rapidly decreasing $\pi$--pole
amplitude. The lower part of Figure~\ref{Figure3} shows
that the transverse cross section can be readily
explained~\cite{Kaskulov:2008xc} by a contribution from DIS pions 
(solid histograms). In the present paper the model of 
Ref.~\cite{Kaskulov:2008xc}
provides an accurate representation of the elementary $(\gamma^*,\pi^+)$ 
cross section.

\begin{figure}[t]
\begin{center}
\includegraphics[clip=true,width=1\columnwidth,angle=0.]
{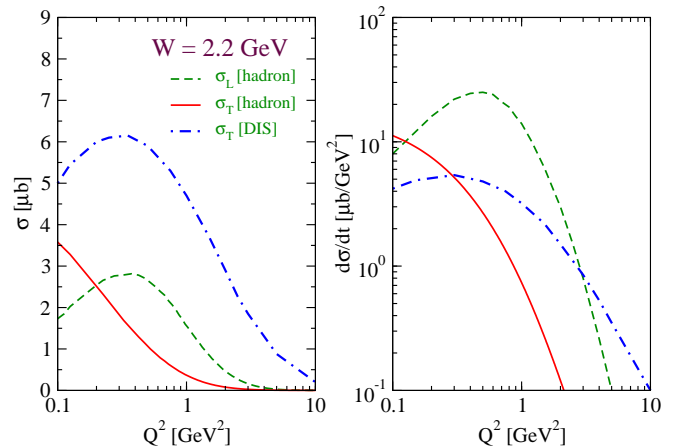}
\caption{\label{Figure4}
Left panel: 
Decomposition of the the integrated virtual photon--nucleon cross sections of the 
exclusive reaction $p(e,e'\pi^+)n$ as a function of the photon
virtuality $Q^2$ at $W=2.2$~GeV. The contribution of the hadron--exchange
model to the transverse $\sigma_{\rm T}$  and longitudinal $\sigma_{\rm L}$
cross sections are shown by  solid and dashed curves, respectively.
The contribution of DIS pions to $\sigma_{\rm T}$ is shown by the dash--dotted 
curve. Right panel: The forward $\pi^+$ differential cross sections as a 
function of $Q^2$. \vspace{-0.8cm}
}
\end{center}
\end{figure}

In Figure~\ref{Figure4} (left panel) we show the integrated
$(\gamma^*,\pi^+)$ cross sections. At given $W=2.2$~GeV the hard partonic part of
the cross section (dash-dotted curve) dominates the $\pi^+$ production
mechanism. However, what matters in the $\pi$CT experiment is the forward
production, since the experiment has been done at parallel kinematics with
$\vec{q}\parallel \vec{k}'$. In this kinematical regime the situation is opposite
and the soft $\pi^+$ quasi--elastic knockout mechanism (dashed and solid curves
correspond to the longitudinal and transverse components, respectively)
dominates up to $Q^2 \approx 3$~GeV$^2$, as shown in the
right panel of Figure~\ref{Figure4}. As we shall see, this complex
interplay between soft hadronic and hard partonic components of the
$(\gamma^*,\pi^+)$ reaction is crucial for the interpretation of $\pi$CT data.

Concerning the overall reaction mechanisms on \emph{nuclei} we rely on a
separation of different processes. At first  (in the impulse approximation)
the $e$--beam interacts with a nucleon inside the nucleus. It is supposed that
the elementary interaction with nucleon is the same as that with a free
nucleon. All the standard nuclear effects like Fermi motion, Pauli blocking
and nuclear shadowing are properly taken into account. In a second step,
all produced (pre)hadrons are propagated through the nuclear medium according to
the transport equation.

\begin{figure}[b]
\begin{center}
\includegraphics[clip=true,width=0.97\columnwidth,angle=0.]
{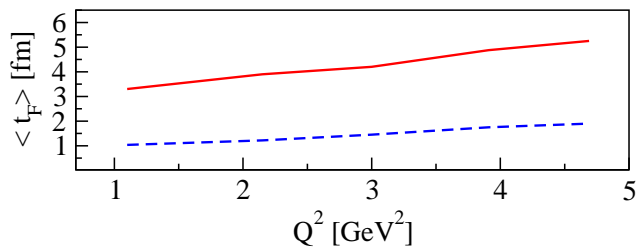}
\caption{\label{Figure5}
The Lund model average formation time of (pre)pions (solid curve) 
in the target rest frame as a function of $Q^2$
in the forward kinematics of the $\pi$CT experiment. 
The dashed curve is an estimate based on Eq.~(\ref{FarrarTime}) with
$\Delta M^2 = 1$~GeV$^2$.
\vspace{-0.4cm}
}
\end{center}
\end{figure}

A necessary condition for the CT effect is the
propagation of a quark--gluon system, originating in the hard partonic
interaction, through the nuclear medium and its subsequent interactions with
surrounding nucleons. In the present model the hard DIS part of the 
primary high energy electromagnetic
interaction is determined by the Lund model which means that the final state
consists of an excited string (see Figure\ \ref{Figure2}). This string then
fragments into hadrons. Following Ref.~\cite{Gallmeister:2005ad}
we extract {\it production} and {\it formation} times $t_P$ and $t_F$ of
hadrons in the
target rest frame 
from the MC calculation. We note however, that in the fragmentation of a string into two
particles all parameters are fixed and the times may be extracted
analytically. 
%The result for {\it formation} time 
%$t_F$ when the (pre)pions evolve to physical states reads
%\begin{eqnarray}
%t_{F} = 
%\frac{2(M_N^2-m_{\pi}^2)\sqrt{\nu^2+Q^2}}{\kappa W^2} + \frac{M_N+\nu}{2
%  \kappa W^2} \hspace{2cm}
% \\
%\times 
%\left[ W^2+\sqrt{16(M_N^2-m_{\pi}^2)^2-8 (M_N^2+m_{\pi}^2) W^2 + W^4} \right]
%\nonumber
%\end{eqnarray}
%where $W^2=M_N^2-Q^2+2\nu M_N$ and $\kappa$=1~GeV/fm is a string tension.
In the exclusive reaction $(e,e'\pi^+)$
considered here all pions are -- because of  their high energy $z \approx 1$
-- directly connected to the hard interaction point and thus have production
time $t_P = 0$.
The Lund model formation times of exclusive (pre)pions in the forward kinematics of the $\pi$CT
experiment are shown in Figure~\ref{Figure5} (solid curve) and include the
dilatation effect in the target rest frame; they are close to zero in the
hadron's rest frame. 
At the highest $Q^2$ the corresponding formation lengths in the laboratory
(the rest frame of the target nucleus)
exceed the nuclear radius so that this effect alone already leads to an
increase  of transparency with $Q^2$.

As a second possible scenario we compare the Lund model {\it formation} time
$t_F$ in the laboratory with
the estimate used in Ref.~\cite{Larson06ge}. In this work
 the characteristic time $t_F$ 
needed to evolve the (pre)pion to its physical state is given by~\cite{farrar}
\begin{equation}
\label{FarrarTime}
t_F \simeq \frac{1}{\sqrt{m_{\pi}^{*2}+(\vec{k}')^2}-\sqrt{m_{\pi}^2+(\vec{k}')^2}}
\approx \frac{2 |\vec{k}'|}{m_{\pi}^{*2}-m_{\pi}^2}.
\end{equation}
where $\vec{k}'$ is the three momentum of the outgoing pion and 
the last approximation is valid only for the ultra-relativistic hadrons.
In the rest frame of the (pre)hadron, {\it i.e.} $\vec{k}'=0$  one gets a well known result
$\tau_F=1/\Delta M$ where $\Delta M=m_{\pi}^{*}-m_{\pi}$ is given by the lowest
lying Regge partner of mass $m_{\pi}^{*}$. Although, in the original work~\cite{farrar}
a rather small value of $\Delta M^2 \simeq 0.25$~GeV$^2$ has been suggested, in Ref.~\cite{Larson06ge}
two values were considered $\Delta M^2 =
1.4$~GeV$^2$ and  $\Delta M^2 = 0.7$~GeV$^2$. 
Clearly, as pointed out in \cite{farrar} all the above estimates 
can be considered as educated guesses at best.
For instance, the value of $\Delta M^2 =
1.4$~GeV$^2$ corresponds
to the expansion time in the rest frame $\tau_F = 0.16$~fm.
This has to be compared with the Lund model which gives the following estimate for 
the {\it formation} time $\tau_F$ in the rest frame of the (pre)pion
\begin{equation}
\tau_F \simeq \frac{m_{\pi}}{\kappa} \frac{1}{z} \approx 0.14~{\rm fm},
\end{equation}
where $z\approx 1$ and a string tension $\kappa=1$~GeV/fm. A direct extraction
of $\tau_F$ from MC~\cite{Gallmeister:2005ad} gives $\tau_F=0.17$~fm.
However, the corresponding times in the laboratory $t_F$ are very different. As one
can see in Figure \ref{Figure5} in the Lund model the {\it formation} time $t_F$ (solid curve) is by about factor of 3 
bigger than $t_F$  obtained when using Eq.~(\ref{FarrarTime}) (dashed
curve). The reason for such differences is related to the fact that in the
Lund model the decay products of the excited strings are projected onto the
states with physical masses and a notion of excited (pre)hadrons
with large masses which then turn in time into the physical ones is not
realized in this scheme. In particular, what is essentially time dilated in the Lund scheme
is the (pre)hadron with the physical mass of the hadron.

\begin{table}[t]
\begin{center}
%\begin{ruledtabular}
\begin{tabular}{|c|c|c|c|c|c|c|c|c|c|}\hline
 $Q^2$ & $\nu$ & $W$ &$-t$ & $\theta_{\gamma}$ & $|\vec{k}'|$ & $|\vec{k}|$ &
 $\varepsilon$ & $x_{\rm B}$& $l_h$\\
 GeV$^2$ & GeV & GeV  & GeV$^2$ & deg & GeV & GeV & &  & fm \\\hline
1.10 & 2.831 &2.26 &0.050  &10.58 &2.793 & 0.23 & 0.50 & 0.21 & 0.67\\
2.15 & 3.282 &2.21 &0.158  &13.44 &3.187 & 0.41 & 0.56 & 0.35 & 0.49\\
3.00 & 3.582 &2.14 &0.289  &12.74 &3.418 & 0.56 & 0.45 & 0.45 & 0.41\\
3.91 & 4.344 &2.26 &0.413  &11.53 &4.077 & 0.70 & 0.39 & 0.50 & 0.39\\
4.69 & 4.733 &2.25 &0.527  & 9.09 &4.412 & 0.79 & 0.26 & 0.54 & 0.36\\\hline
\end{tabular}
\caption{\label{table1} \small The central kinematics of the $\pi$CT
 experiment in the laboratory~\cite{Clasie:2006re}. Here $t$ and
$|\vec{k}|$ stand for the four and three momentum transfer to the 
target, respectively, $\theta_{\gamma}$ is the angle between the three 
momentum of the virtual photon and the electron beam direction in the 
laboratory, $\varepsilon$ is the virtual photon polarization, the Bjorken 
variable $x_{\rm B}=Q^2/2M_N\nu$ and $l_h$ denotes the coherence length 
for each kinematics settings. For the parallel kinematics $\theta_{\gamma} 
= \theta_{\pi}$ where $\theta_{\pi}$ is the angle of the emitted pion 
with three momentum $|\vec{k}'|$. \vspace{-0.7cm}} 
\end{center}
\end{table}

\begin{figure*}[t]
\begin{center}
\includegraphics[clip=true,width=1.45\columnwidth,angle=0.]
{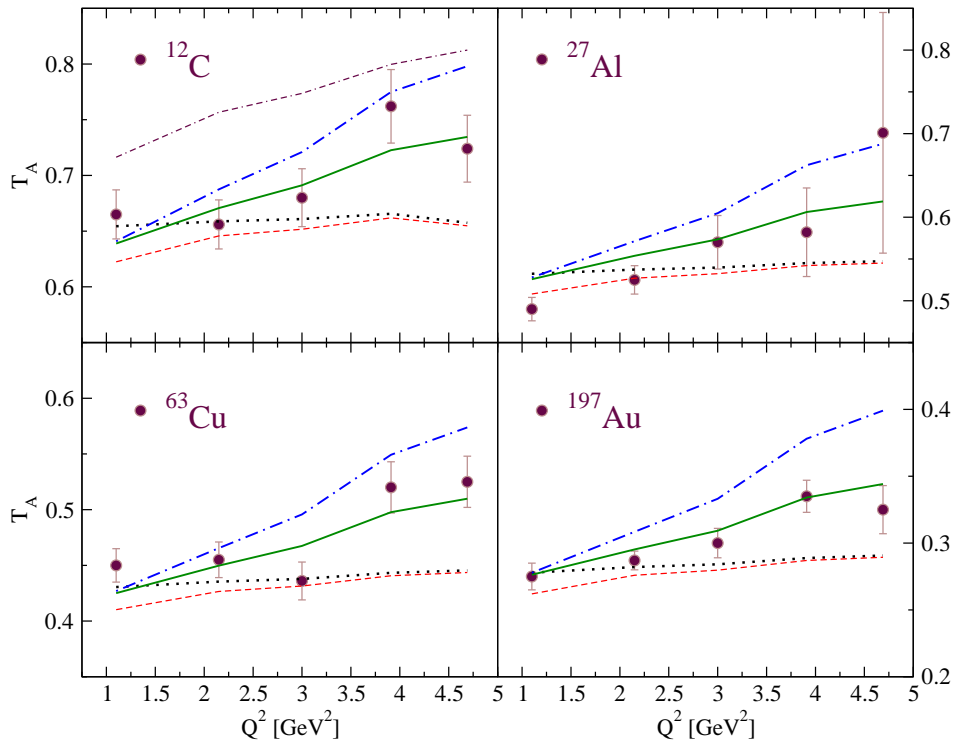}
\caption{\label{Figure7}
\small  Transparency, $T_{A}$, vs. $Q^2$ for $^{12}$C (left, top panel), $^{27}$Al (right, top), $^{63}$Cu (left, bottom) and  $^{197}$Au (right,
bottom). 
The dotted curves correspond to FSI with the full hadronic cross section and the dashed curves include
the shadowing corrections. 
The dash--dotted curves correspond to the in--medium
cross sections defined according to the Lund model formation time concept which
includes the $Q^2$--dependent (pre)hadronic interactions,
Eq.~(\ref{eq:scenarioQ}), for the transverse contribution. The solid curves
describe the effect of time dilatation alone with the pedestal value in the
effective cross section  independent
of $Q^2$. The dash--dash--dotted curve in the top left panel realises the CT
effect both in the longitudinal and transverse channels. 
The experimental data are from Ref.~\cite{:2007gqa}. \vspace{-0.6cm}} 
\end{center}
\end{figure*}

\begin{figure*}[t]
\begin{center}
\includegraphics[clip=true,width=1.45\columnwidth,angle=0.]
{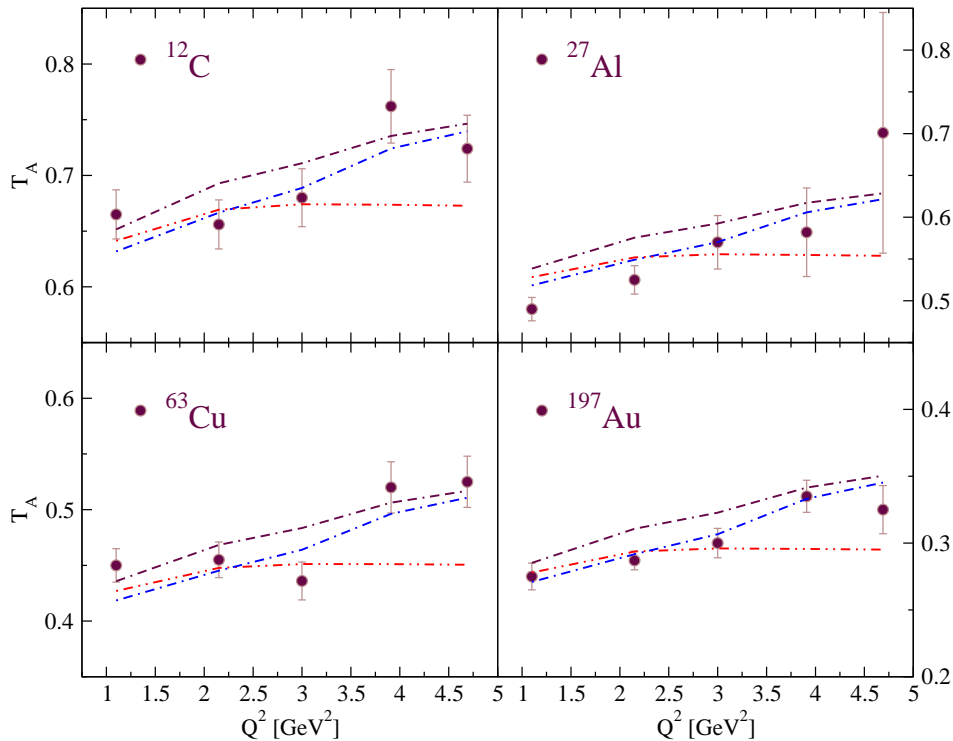}
\caption{\label{Figure8}
\small  Transparency, $T_{A}$, vs. $Q^2$ for $^{12}$C (left, top panel), $^{27}$Al (right, top), $^{63}$Cu (left, bottom) and  $^{197}$Au (right,
bottom). The formation time of (pre)pions in the laboratory is calculated using Eq.~(\ref{FarrarTime}).
The dash--dash--dotted curves realize the CT effect in both the
longitudinal
and transverse channels and dash--dotted curves in the transverse channel only. 
The dot--dot--dashed curves describe the CT effect in the longitudinal
channel only.
%The notations for the curves are the same as in Figure~\ref{Figure7}.
%The dotted curves correspond to FSI with the full hadronic cross section and the dashed curves include
%the shadowing corrections. The solid curves account for the formation time concept.
The experimental data are from Ref.~\cite{:2007gqa}. 
\vspace{-0.6cm} }
\end{center}
\end{figure*}

We further use the quantum diffusion model of Ref.~\cite{farrar} to describe the
time--development of the interactions of a point--like configuration
produced in a hard initial reaction. This approach combines a linear
increase of the hadron--nucleon cross section with the assumption that the
cross section for the leading particles does not start at zero, but at a
finite pedestal value connected with $Q^2$ of the initial interaction, {\it i.e.}
\begin{equation}
\label{effcs}
\sigma^*(t)/\sigma  = X_0
+(1-X_0) \left({(t-t_P)}/{(t_F-t_P)}\right),
\label{eq:scenarioQ}
\end{equation}
where $X_0 = {r_{\rm lead}}\frac{const}{Q^2}$ with $r_{\rm lead}$ standing for
the ratios of leading partons over the total number of partons (2 for mesons,
3 for baryons). The baseline value $X_0$ is inspired by the coefficient $n^2 \langle
k_{\rm t}^2\rangle/Q^2$ in \cite{farrar}, where $n=2$ for mesons and $\langle k_{\rm
  t}^2\rangle$ denotes the average transverse momentum of partons.
The scaling with $r_{\rm lead}$ guarantees that summing over all particles in
an event, on average the prefactor becomes unity.
The size of the pion enters through the (pre)hadronic cross sections 
 with the pedestal $1/Q^2$ behavior
and also through the linear rise of the (pre)hadronic interaction 
cross section.
Using Eq.~(\ref{effcs}) the (pre)hadronic cross section is zero before 
the {\it production} time $t_P$ and
equals the full hadronic cross section after the {\it formation} time $t_F$. 
We note that, following the Lund model hadronization pattern this
(pre)hadronic interaction is effective only for the DIS events;  the
longitudinal cross section is not affected by this (pre)hadronic
interaction. Thus, in the model of Ref.~\cite{Kaskulov:2008xc} 
only the DIS (transverse) part of the cross section is
responsible for the observed effect since this part is connected
with the 4D pattern of the string breaking dynamics which makes the
formation time of (pre)hadrons finite.

The propagation of the produced (pre)hadron through the nuclear medium is
described by the Boltzmann--Uehling--Uhlenbeck (BUU) equation
which describes the time evolution of the phase space density $f_i(\vec r,\vec
p,t)$ of particles of type $i$ that can interact via binary reactions.
Besides the produced hadron and the nucleons these particles involve the baryonic resonances and mesons
that can be produced in FSI. For the baryons the equation contains a mean
field potential which depends on the particle position and momentum. The BUU
equations of each particle species $i$ are coupled via the mean field and the
collision integral. The latter allows for elastic and inelastic
rescattering and side--feeding through coupled--channel effects; it accounts
for the creation and annihilation of particles of type $i$ in a secondary
collisions as well as elastic scattering from one position in phase space into
another. The resulting system of coupled differential--integral equations
is solved via a test particle ansatz for the phase space density. For fermions
Pauli blocking is taken into account via blocking factors in the collision 
term~\cite{GiBUU}. We note that exactly this method leads to a good
simultaneous  description of hadronic attenuation in nuclei observed both by
the  EMC (200 GeV) and the HERMES (27 GeV)
experiments~\cite{Gallmeister:2007an}. The model also works very well 
for semi--exclusive reactions as has been shown, for example, in analysis
of the photoproduction of mesons on 
nuclei~\cite{Krusche:2004zc,Mertens:2008np}. 
In \cite{Krusche:2004zc} it has been shown, that for
pions coherent production, which is outside the applicability of semiclassical
transport, does not play a role above the $\Delta(1232)$--resonance.

Following Ref.~\cite{:2007gqa} the nuclear transparency ratio in the reaction
$A(e,e'\pi^+)$ is defined as
\begin{equation}
\label{Tbuu}
T_A = {\sigma}_A^{\rm FSI}/\sigma_A,
\end{equation}
where the cross sections $\sigma_A$ (in the laboratory) read
\begin{equation}
\label{sigmaTbuu}
\sigma_A = \int_{\Delta M_X} d M_X \frac{d\sigma_A}{dQ^2 d\nu d\Omega_{\pi} dM_X}.
\end{equation}
In Eq.~(\ref{Tbuu}) $\sigma_A^{\rm FSI}$ and $\sigma_A$ are the results of the
model calculations  with and without FSI of hadrons in their way out of the
nucleus, respectively. In Eq.~(\ref{sigmaTbuu}) $M_X$ stands for the missing
mass of the recoiling nuclear system. It can be calculated using the four
momenta of the virtual photon $q$ and the four momentum  of the detected pion $k'$:
$M^2_X = (q+ P_{A} - k')^2$, where $P_{A}=(M_A,\vec{0})$ denotes the four momentum
of the nuclear target in the laboratory. In Eq.~(\ref{sigmaTbuu}) $\Delta M_X$
denotes the region of the missing mass $M_X$ within the experimental
acceptance. In the case of the free proton target, the missing mass is a
$\delta$--function at the neutron mass. For nuclei, the Fermi motion of the
bound nucleons broadens the distributions, and the missing mass is limited by
the acceptance conditions. The lower limit of $M_X$ is fixed by the
$1\pi$ production threshold and the upper limit of $M_X$ in the integration is
determined by the values imposed in the actual experiment. The latter is done
to reduce the contamination from multi--pion events in the final
yield. This  guarantees that in the simulation procedure, like in the actual
experiment, one selects events containing only single $\pi^+$ and the residual
excited nucleus. The positions of the above--threshold nuclear missing mass
cuts for all the kinematic settings (see Table \ref{table1}) and targets are 
taken from Ref.~\cite{Clasie:2006re} and used in the calculations.

Note that the shadowing (ISI) of the virtual photon is rather small for the
kinematic conditions of the $\pi$CT experiment and varies
weakly with $Q^2$. The coherence length $l_h\simeq 2\nu/Q^2+m_{h}$, where
$m_h$ stands for the mass of the $\rho$--meson, varies (see Table \ref{table1})
from $\simeq 0.67$~fm at $Q^2=1.1$~GeV$^2$ down to $0.36$~fm, {\it i.e.}\ less than
one nucleon's radius, at $Q^2=4.69$~GeV$^2$, only.
The ISI effect for the VMD like $\pi^+$ quasi--elastic knockout part is
included in the numerator  of Eq.~(\ref{Tbuu}) using the method of 
Ref.~\cite{Falter:2002vr}.

We have performed the calculations for the nuclei $^{12}$C,
$^{27}$Al, $^{64}$Cu and $^{197}$Au at identical kinematics shown in
Table \ref{table1}. Here the central values correspond to the parallel 
kinematics ($\theta_{\gamma} = \theta_{\pi}$) where the magnitude of the
outgoing pion three momentum $|\vec{k}'|$ is given by
\begin{equation}
|\vec{k}'| = \sqrt{Q^2+\nu^2}-|\vec{k}|,
\end{equation}
where $|\vec{k}|$ is the three momentum transfer to the nucleus. The parallel
kinematics $\vec{q}\parallel \vec{k}'$ is supposed to minimize the
contribution of the elastic rescattering~\cite{Strikman:2007nv}.

In Figure~\ref{Figure7} we show our results with different scenarios for the hadronic
FSI. At first, we consider a case where all the produced hadrons interact from
their production on with the full hadronic cross sections.  The result of this 
calculation is shown by the dotted curves. These curves are nearly flat and
thus do not exhibit the observed rise with $Q^2$. The same is true for the 
short--dashed curves which include in addition the ISI shadowing; the
influence of the latter is very small due to the very small coherence lengths 
as discussed above.

The dash--dotted curves in Figure~\ref{Figure7} correspond to the in--medium
cross sections defined according to the Lund model formation time concept which
includes the $Q^2$--dependent (pre)hadronic interactions,
Eq.~(\ref{eq:scenarioQ}), for the transverse contribution.
Here the pedestal value $X_0$ has been fixed by assuming $\sqrt{\langle
  k_{\rm t}^2\rangle}=350$~MeV, i.e.\ the value used in the studies 
of~\cite{Larson06ge,Cosyn:2007er}. This model reproduces the trend of the data 
but on average overestimates the transparency for all four nuclear targets.
The increase of $T_A$, which we see at high $Q^2$ in this scenario, is driven 
by the $(t-t_F)$ factor in Eq.~(\ref{eq:scenarioQ}) and
the time dilatation effect seen in Figure~\ref{Figure5}. The latter
results in an increase of $t_F$ as a 
function of $Q^2$ in the target rest frame or, equivalently, the $\pi^+$ three momentum. 

To show the effect of time dilatation alone we take in
Eq.~(\ref{eq:scenarioQ}) the pedestal value $X_0=r_{\rm lead}=1/2$ independent
of $Q^2$. The result of this assumption is shown by the solid curves and provides
a good agreement with data. However, there is an interplay
between the formation times used and the $Q^2$--dependent pedestal value
in the effective cross section, see Eq.~(\ref{effcs}). Either a larger pedestal value or a decreased
$t_F$
would lower the dash--dotted line in Figure~\ref{Figure7} toward the data.
We note that the scenario leading to the solid curve (no $Q^2$ dependence)
is different from that in Refs.~\cite{Larson06ge,Cosyn:2007er} where
a $Q^2$--dependent pedestal value had been used.

The calculations of Refs.~\cite{Larson06ge,Cosyn:2007er} show a pronounced CT 
effect already at values of $Q^2$ as low as $1.1$~GeV$^2$. In these works it 
was assumed that {\it all} produced $\pi^+$ are subject to CT, whereas in our 
model the $\pi$--pole mechanism, which at low $Q^2$ gives the dominant
contribution to the forward production of $\pi^+$, is not affected by the
formation time effect. To model this situation we assume the same $t_F$ 
for all pions produced by the longitudinal and transverse photons. In Figure~\ref{Figure7} 
(left, top panel) the result of such a calculation is shown by the
dash--dash--dotted curve; it is seen to strongly overestimate the transparency 
ratio. Thus, assuming a finite $t_F$ also for preexisting pions which are
knocked out from the nucleon meson cloud ($t$--channel process) will
destroy the agreement obtained above. 
The fact that the calculations of Refs.~\cite{Larson06ge,Cosyn:2007er} 
do fit the data is due to the facts that there 1)
the {\it formation} times of (pre)pions in the laboratory, 
essentially free parameters in~\cite{Larson06ge,Cosyn:2007er}, 
are smaller
than those extracted from the Lund model and 2) 
in Ref.~\cite{Larson06ge} the total cross section was 
used in the Glauber attenuation formula. The latter may be a good approximation for 
the strictly forward kinematics in this experiment 
where any elastic 
process would scatter the pion out of the 
forward acceptance. However, because of the finite experimental 
resolution and the 
acceptance cuts~\cite{Clasie:2006re} % $\simeq \pm 8~\%$~\cite{Clasie:2006re} 
around the central
values  of the pion three momentum (see Table~\ref{table1}),
the ideal forward kinematics is not realized in the $\pi$CT experiment.
As a result the attenuation in the $\pi$CT experiment is not driven necessarily by the
total $\pi^+ N$ cross section.
%Furthermore, the Glauber approach neglects the effect of
%Pauli--blocking which is most effective for forward scattering. 

So far we have considered the (pre)hadronic expansion times extracted from
the string breaking pattern of the Lund model. In Figure~\ref{Figure8} we
present the results with $t_F$ calculated when using Eq.~(\ref{FarrarTime}) --
the concept realized in Refs.~\cite{Larson06ge,Cosyn:2007er}. The
calculations were done for $\Delta M=1$~GeV as a fit parameter. 
This is an optimal value needed to reproduce the $\pi$CT data with our
treatment of FSI.
The dash--dash--dotted curves realize the CT effect in both the
longitudinal
and transverse channels and dash--dotted curves in the transverse channel only.
In addition we show the results of the CT effect in the longitudinal
channel only (dot--dot--dashed curves). As one can see the latter scenario
is certainly ruled out by the present data. Because of the dominance of the
transverse cross section at high values of $Q^2$, a use of different values of $\Delta
M$ in a range discussed before does not change this result significantly. This is 
particularly interesting because presently the CT effect is expected to show
up in the longitudinal channel~\cite{Strikman:2007nv}.

%Contrary to the work of~\cite{Larson06ge,Cosyn:2007er} 
Our results based on
the Lund model hadronization scheme and {\textsc
l}--{\textsc t} separated transparencies presented in Figure~\ref{Figure8}
suggest that CT shows up in the transverse part of the
$\gamma^*$--nucleus cross section $\sigma_{\rm T}$. It would, therefore, 
be interesting to see the $Q^2$ dependence of {\textsc
l}--{\textsc t} separated experimental  cross sections for the
transparency. The ongoing experiments at JLAB~\cite{Dutta} may verify 
this conclusion.

In summary, in this work we have presented a calculation of the nuclear 
transparency of pions in the reaction $A(e,e'\pi^+)A^*$ off nuclei. The
microscopic input for the primary interaction of the virtual photon with 
the nucleon describes both the transverse and the longitudinal cross sections.
The coupled--channel BUU transport model has been used to describe the FSI of
hadrons in the nuclear medium. The formation times of (pre)hadrons follow the 
time--dependent hadronization pattern of hard DIS processes.
Our results are consistent with the JLAB data and show that a detailed
understanding of the primary $\gamma^*N$ interaction may be essential for a
quantitative understanding (and proof) of CT. The {\textsc
l}--{\textsc t} separated cross sections would be extremely useful for this
aim. It would, furthermore, be interesting to extend the present analysis 
to $Q^2 \sim $ 10~GeV$^2$, where the largest CT effects are predicted.

We gratefully acknowledge helpful communications and discussions 
with D. Dutta, R. Ent, H. Gao and G. Miller. 
We appreciate helpful discussions with
O. Buss, T. Gaitanos and the GiBUU group.

This work was supported by BMBF.

\end{document}